# Impact of Top SiO$_2$ Interlayer Thickness on Memory Window of Si Channel FeFET with TiN/SiO$_2$/Hf$_{0.5}$Zr$_{0.5}$O$_2$/SiO$_x$/Si (MIFIS) Gate Structure

Tao Hu, Xianzhou Shao, Mingkai Bai, Xinpei Jia, Saifei Dai, Xiaoqing Sun, Runhao Han, Jia Yang, Xiaoyu Ke, Fengbin Tian, Shuai Yang, Junshuai Chai, Hao Xu, Xiaolei Wang, Wenwu Wang, and Tianchun Ye

*Abstract*—**We study the impact of top SiO$_2$ interlayer thickness on the memory window (MW) of Si channel ferroelectric field-effect transistor (FeFET) with TiN/SiO$_2$/Hf$_{0.5}$Zr$_{0.5}$O$_2$/SiO$_x$/Si (MIFIS) gate structure. We find that the MW increases with the increasing thickness of the top SiO$_2$ interlayer, and such an increase exhibits a two-stage linear dependence. The physical origin is the presence of the different interfacial charges trapped at the top SiO$_2$/Hf$_{0.5}$Zr$_{0.5}$O$_2$ interface. Moreover, we investigate the dependence of endurance characteristics on initial MW. We find that the endurance characteristic degrades with increasing the initial MW. By inserting a 3.4 nm SiO$_2$ dielectric interlayer between the gate metal TiN and the ferroelectric Hf$_{0.5}$Zr$_{0.5}$O$_2$, we achieve a MW of 6.3 V and retention over 10 years. Our work is helpful in the device design of FeFET.**

## I. Introduction

Hafnia(HfO$_2$) based silicon channel ferroelectric field-effect transistors (HfO$_2$ Si-FeFETs) have been extensively studied, as a strong candidate for non-volatile memory with lower power operation, fast speed, and CMOS compatibility [1-13], thanks to the discovery of ferroelectricity in doped-HfO$_2$ [14]. The large spontaneous polarization ($P_s$) of ferroelectric doped-HfO$_2$ (~ 2-30 μC/cm$^2$) induces significant charge trapping and de-trapping phenomenon, and this results in the decrease of the memory window (MW) [15-18]. To suppress the charge trapping and de-trapping effect, several studies focus on suppressing the charge injection from the silicon channel to the ferroelectric Hf$_{0.5}$Zr$_{0.5}$O$_2$/Si interface, such as applying high-κ interlayer [19-23], reducing the spontaneous polarization of ferroelectric [24-27], and eliminating interlayer [28-32]. However, the above method cannot significantly improve the MW and the MW is generally limited to less than 2 V. This MW does not meet the requirements for application in multi-bit memory cells. Recently, inserting a dielectric layer (e.g. SiO$_2$ or Al$_2$O$_3$) between the metal gate and ferroelectric layer was found to be an effective method to significantly improve the MW [33-38]. The MW can achieve 3.8 V by inserting 2 nm SiO$_2$ and 4.1 V by inserting 3 nm Al$_2$O$_3$ [34, 38]. Moreover, the simulation results show that using a low dielectric constant and thick interlayer (such as SiO$_2$) is beneficial for the MW increase [35].

However, there are still no experimental studies on this issue and the impact of the top SiO$_2$ interlayer thickness on the MW is still unclear. Therefore, we experimentally report the dependence of the MW on the top SiO$_2$ interlayer thickness in this work and discuss its physical origin. The physical origin of the impact of the top SiO$_2$ thickness on the MW is attributed to two discrete defect energy levels. We find that the MW increases with the increasing thickness of the top SiO$_2$ interlayer. We achieve a maximum MW of 6.3 V by inserting a 3.4 nm top SiO$_2$ interlayer. Furthermore, we study the dependence of the endurance characteristics on initial MW. We find that the endurance characteristic degrades with increasing the initial MW. We realize an endurance of ~ 10$^4$ cycles with an initial MW of 5.3 V by inserting a 3.4 nm top SiO$_2$ interlayer.

## II. Device Fabrication And Characterization

Fig. 1 shows the schematic of the device structure and fabrication process flow. In this work, there are two different gate stacks. One is TiN/Hf$_{0.5}$Zr$_{0.5}$O$_2$/SiO$_x$/Si (MFIS) as the control sample. The other is TiN/SiO$_2$/Hf$_{0.5}$Zr$_{0.5}$O$_2$/SiO$_x$/Si (MIFIS) with a 0.85, 1.7, 2.55, or 3.4 nm top SiO$_2$ interlayer.

These devices were fabricated in an 8-inch P-type silicon wafer with a resistivity of 8-12 Ω·cm using a gate-last process. Firstly, B ions with an energy of 60 KeV and a dose of 8 × 10$^{12}$ cm$^{-2}$ were implanted into the whole Si substrate. After that, the Si substrate was annealed at 1050 °C for 4 h and 1150 °C for 40 min to form a P-type well. Next, the source and drain regions

This work was supported by the National Natural Science Foundation of China under Grant No. 92264104 and 52350195 and the Postdoctoral Fellowship Program of CPSF under Grant No. GZC20232925. (Corresponding author: Xiaolei Wang)

Tao Hu, Xianzhou Shao, Mingkai Bai, Xinpei Jia, Saifei Dai, Xiaoqing Sun, Runhao Han, Jia Yang, Xiaoyu Ke, Fengbin Tian, Shuai Yang, Junshuai Chai, Hao Xu, Xiaolei Wang, Wenwu Wang, and Tianchun Ye are with Institute of microelectronics of the Chinese Academy of Sciences, Beijing 100029, China. The authors are also with the School of Integrated Circuits, University of Chinese Academy of Sciences, Beijing 100049, China (e-mail: wangxiaolei@ime.ac.cn).

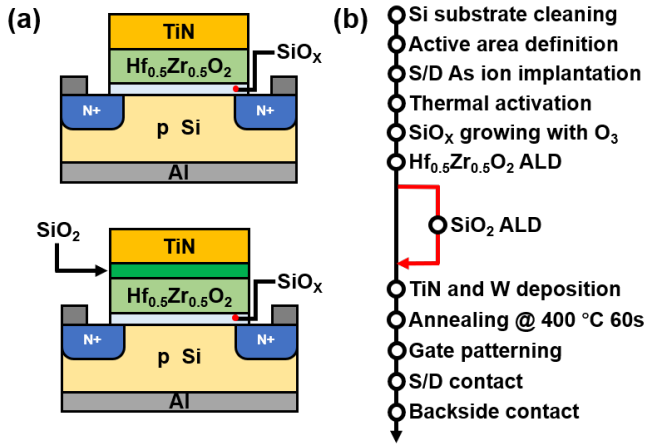

Fig. 1. (a) Schematic of the HfO$_2$ Si-FeFET device structure and (b) fabrication process flow.

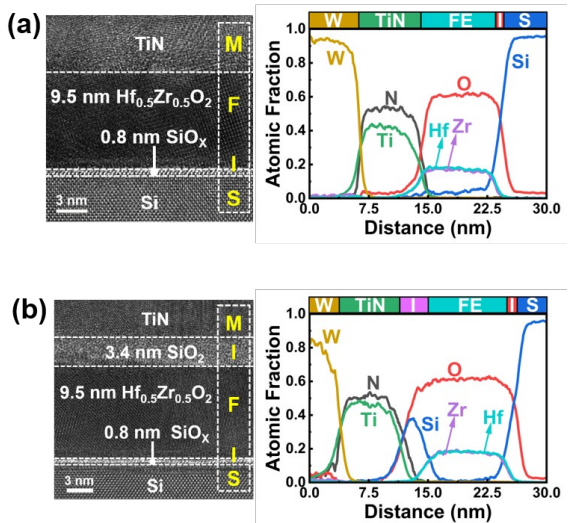

Fig. 2. HRTEM images and EDS of the (a) MFIS and (b) MIFIS structures.

were defined through photolithography. Then As ions with an energy of 50 KeV and a dose of $2 \times 10^{15}$ cm$^{-2}$ were implanted into source and drain regions. After that, these devices were annealed at 1050 °C for 5 s under the N$_2$ atmosphere for dopant activation. Subsequently, the gate stack was formed. After diluted-HF clean, the 0.8 nm botten SiO$_x$ interlayer was grown by ozone oxidation at 300 °C. Then, 9.5 nm Hf$_{0.5}$Zr$_{0.5}$O$_2$ and 0.85-3.4 nm top SiO$_2$ interlayer were grown by atomic layer deposition (ALD) at 300 °C. For the growth of ALD Hf$_{0.5}$Zr$_{0.5}$O$_2$, the tetrakis-(ethylmethylamino)-hafnium (TEMA-Hf), tetrakis-(ethylmethylamino)-zirconium (TEMA-Zr), and H$_2$O were used as precursors of Hf, Zr, and O, respectively. For the growth of ALD SiO$_2$, the bis(diethylamino)silane (BDEAS) and ozone were used as precursors of Si and O, respectively. Then, 10 nm TiN was grown by physical vapor deposition (PVD) and 75 nm W was grown by ALD. The devices were annealed at 400 °C under an N$_2$ atmosphere for 60 s by using rapid thermal annealing (RTA) to form the orthorhombic phase. Finally, the forming gas annealing (FGA) was performed at 450 °C in 5%-H$_2$/95%-N$_2$ to passivate the defects of the gate stack. All of the fabrication processes of the MFIS structure were the same as the MIFIS structure, except without the top dielectric interlayer SiO$_2$.

Fig. 2 shows High-Resolution Transmission Electron Microscopy (HRTEM) images and Energy Dispersion Spectrometer (EDS) results for both MFIS and MIFIS structures. For the MIFIS structure, the presence of a peak concentration of Si at the TiN/Hf$_{0.5}$Zr$_{0.5}$O$_2$ interface confirms the presence of the top SiO$_2$ interlayer.

In this work, the gate length/width (L/W) of these devices is 5/150 μm. The electrical measures were performed by Keysight B1500A with a waveform generator fast measurement unit (WGFMU) and high voltage semiconductor pulse generator unit (SPGU). The threshold voltage ($V_{th}$) is extracted by the constant current method at the drain current $I_d$ = W/L × 10$^{-7}$ A.

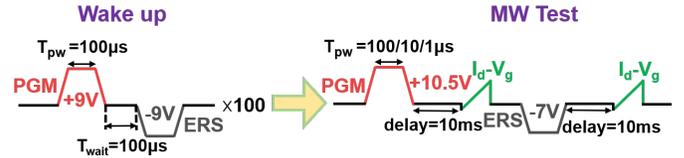

Fig. 3. Waveforms of electrical measurement.

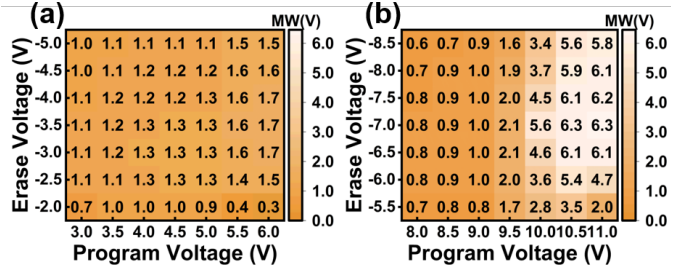

Fig. 4. The MW mapping of (a) MFIS and (b) MIFIS with 3.4 nm top SiO$_2$ as a function of the pulse amplitude under the pulse width of 100 μs.

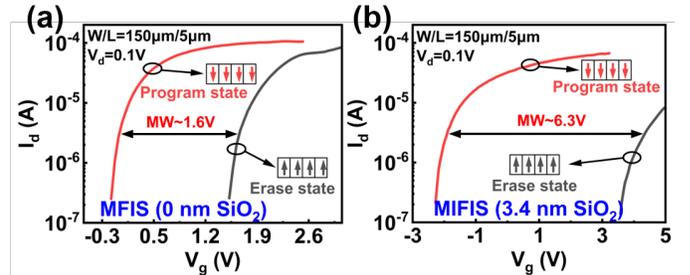

Fig. 5. $I_d$–$V_g$ curves of the maximum MW for (a) MFIS and (b) MIFIS with 3.4 nm top SiO$_2$ interlayer under the pulse width of 100μs.

## III. RESULTS AND DISCUSSIONS

### A. The dependence of MW on the top SiO$_2$ thickness

We investigate the dependence of the MW on the pulse amplitude. Fig. 3 shows the MW measurement waveforms. The pulse width is set as 100 μs. Fig. 4 shows the MW mapping results for the MFIS and MIFIS structure with the 3.4 nm top SiO$_2$ interlayer (unless specified otherwise in the following context, MIFIS structure refers to the MIFIS structure with the 3.4 nm top SiO$_2$ interlayer). For the MFIS or MIFIS structure, when the program pulse amplitude goes beyond 6 V or 11 V under the pulse width of 100 μs, respectively, the devices break down. Thus, we find that the maximum MW is 6.3 V for the MIFIS structure with the 3.4 nm top SiO$_2$ interlayer, while the maximum MW is 1.7 V for the MFIS sample. Fig. 5(a) and (b) show the $I_d$-$V_g$ curves corresponding to the maximum MW for the MFIS and MIFIS structures, respectively.

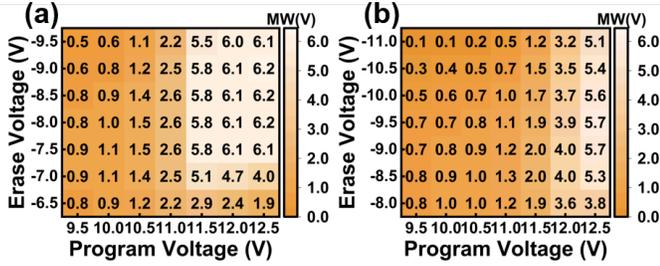

Fig. 6. The MW mapping of MIFIS with 3.4 nm top SiO$_2$ as a function of the pulse amplitude under the pulse width of (a)10 µs and (b) 1 µs.

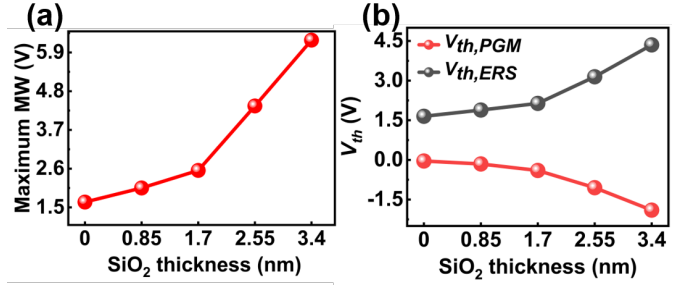

Fig. 7. The dependence of (a) the maximum MW and (b) the $V_{th}$ corresponding to the maximum MW on the top SiO$_2$ interlayer thickness.

We repeated the above measurement process under the different pulse widths. Fig. 6(a) and (b) show the MW mapping results for the MIFIS structure under the pulse width of 10 µs and 1 µs. The results are similar to the pulse width of 100 us in Fig. 4. Thus the pulse width does not affect the conclusion in this work.

We investigate the impact of top SiO$_2$ interlayer thickness on the maximum MW. For each sample, we repeated the above MW mapping measurement process. Fig. 7(a) shows the dependence of the maximum MW on the top SiO$_2$ interlayer thickness. Fig. 7(b) shows the dependence of the $V_{th}$ on the top SiO$_2$ interlayer thickness corresponding to the maximum MW in Fig. 7(a). From Fig. 7, we can see that the maximum MW of the MIFIS structure increases with top SiO$_2$ thickness. Moreover, the curve of the maximum MW vs. the top SiO$_2$ thickness exhibits a two-stage linear relationship.

### B. The physical origin of the MW dependence on the top SiO$_2$ thickness

Firstly, we discuss the physical origin of MW enlargement by inserting the top SiO$_2$ interlayer compared with the sample without the top SiO$_2$ interlayer. The origin is charge trapping between the TiN metal gate and the top SiO$_2$/Hf$_{0.5}$Zr$_{0.5}$O$_2$ interface [34, 35, 38]. During the erase pulse, electrons trapping and/or holes de-trapping at the top SiO$_2$/Hf$_{0.5}$Zr$_{0.5}$O$_2$ interface occurs. This results in negative charges at the top SiO$_2$/Hf$_{0.5}$Zr$_{0.5}$O$_2$ interface because the top SiO$_2$ acts as a barrier layer after the erase operation. Then the $V_{th}$ positively shifts. During the program pulse, electrons de-trapping and/or holes trapping at the top SiO$_2$/Hf$_{0.5}$Zr$_{0.5}$O$_2$ interface occurs. This results in positive charges at the top SiO$_2$/Hf$_{0.5}$Zr$_{0.5}$O$_2$ interface after the program. Then the $V_{th}$ negatively shifts. Finally, the above results lead to a large MW. Fig. 7(b) supports the above discussion.

Secondly, we discuss the physical origin of the dependence of the maximum MW on the top SiO$_2$ thickness. Due to the presence of the trapped charges ($Q_{tp}$) at the top SiO$_2$/Hf$_{0.5}$Zr$_{0.5}$O$_2$ interface, the shift of the threshold voltage ($\Delta V_{th}$) with respect to the control sample is calculated by

$$\Delta V_{th} = -\frac{Q_{tp}}{\varepsilon_0 \varepsilon_{SiO_2}} \cdot d_{SiO_2} \quad (1)$$

where $\varepsilon_0$ is the vacuum dielectric constant, $\varepsilon_{SiO_2}$ is the relative dielectric constant of the top SiO$_2$ interlayer, and $d_{SiO_2}$ is the physical thickness of the top SiO$_2$ interlayer. When the interfacial charges ($Q_{tp}$) remain constant, the shift of the threshold voltage ($\Delta V_{th}$) is linearly dependent on the top SiO$_2$ thickness. Thus, the results indicate that the two-stage relationship between the maximum MW and the top SiO$_2$ thickness as shown in Fig. 7(a) is due to the presence of different interfacial charges $Q_{tp}$, which increases with an increase in the thickness of the top SiO$_2$.

We use the energy band diagram to discuss the physical origin. Fig. 8(a) shows the band diagram of the MIFIS during the erase operation. We consider that both donor and acceptor traps appear at the top SiO$_2$/Hf$_{0.5}$Zr$_{0.5}$O$_2$ interface. The donor and acceptor traps are located near the valence and conduction band of Hf$_{0.5}$Zr$_{0.5}$O$_2$, respectively. During the erase operation, the electric field across the top SiO$_2$ interlayer is nearly close to its breakdown field due to the large spontaneous polarization $P_s$ (~20 µC/cm$^2$) of the ferroelectric Hf$_{0.5}$Zr$_{0.5}$O$_2$. This results in significant band bending of the top SiO$_2$ interlayer. Thus, the Fermi energy level of the gate metal TiN with the thicker top SiO$_2$ interlayer is higher than that of the thinner top SiO$_2$ interlayer. Then the donor traps are filled by the tunneling electrons regardless of the top SiO$_2$ thickness and show neutral charges. For the SiO$_2$ thickness of less than 1.7 nm, the Fermi energy level of the gate metal TiN is always located above the acceptor trap energy level $E_{a,1}$ during the erase operation. The acceptor trap energy level $E_{a,1}$ is also filled by the electrons and shows negative charges. After the erase operation, these negative charges are left as shown in Fig. 8(b), and lead to the positive shift of $V_{th}$ after the erase operation ($V_{th,\,ERS}$) as shown in Fig. 7(b). Moreover, for the SiO$_2$ thickness of less than 1.7 nm, these negative charges determined by acceptor trap energy level $E_{a,1}$ after the erase operation are nearly identical, and this results in the first stage of $V_{th,\,ERS}$ vs. the top SiO$_2$ thickness curve as shown in Fig. 7(b).

We discuss the physical origin of the second stage of the erase case. The second stage indicates that the acceptor traps have two levels, which is schematically shown in Fig. 8(a). For the SiO$_2$ thickness of less than 1.7 nm, the acceptor trap energy level $E_{a,1}$ is filled by the electrons and the acceptor trap energy level $E_{a,2}$ is empty. For thicker SiO$_2$ thickness of larger than 1.7 nm, however, both the acceptor trap energy level $E_{a,2}$ and $E_{a,1}$ are filled because thicker SiO$_2$ induces a larger voltage drop across the whole top SiO$_2$ interlayer and higher Fermi level of the metal gate with respect to the trap levels as shown in Fig. 8(a). Then more negative charges are trapped for thicker SiO$_2$ thickness of larger than 1.7 nm, and result in the second stage

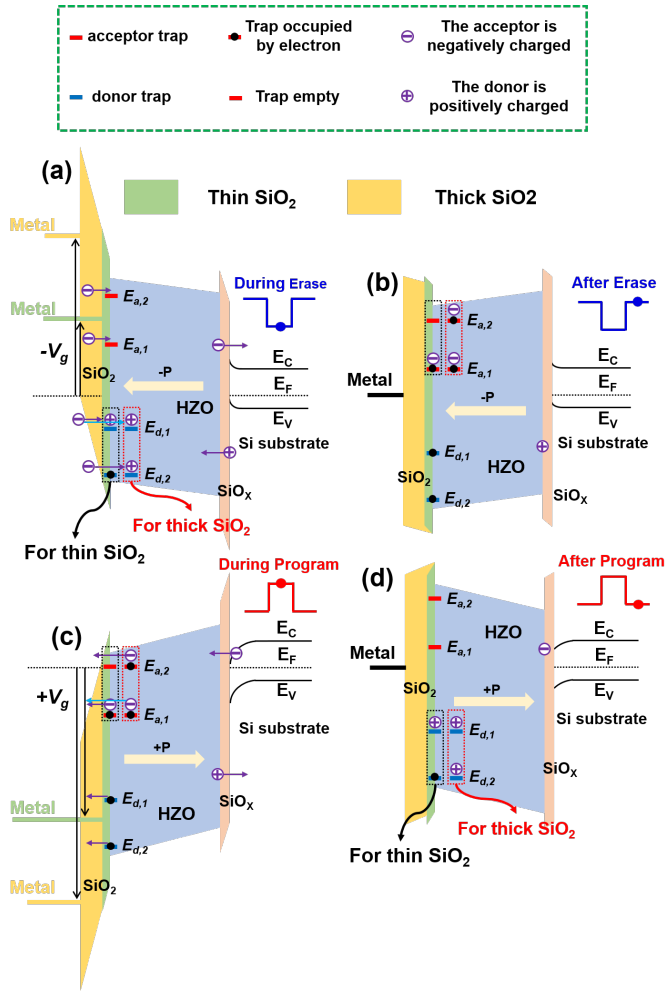

Fig. 8. Band diagram of the MIFIS with different top SiO$_2$ interlayer thickness: (a) during erase, (b) after erase, (c) during program and (d) after erase.

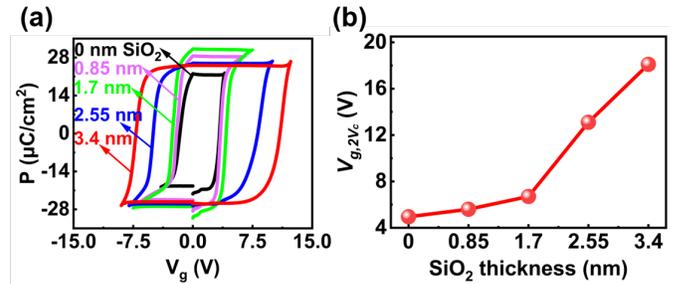

Fig. 9. (a) The results of the PUND measurements. (b) The dependence of $V_{g,2Vc}$ on the top SiO$_2$ interlayer thickness.

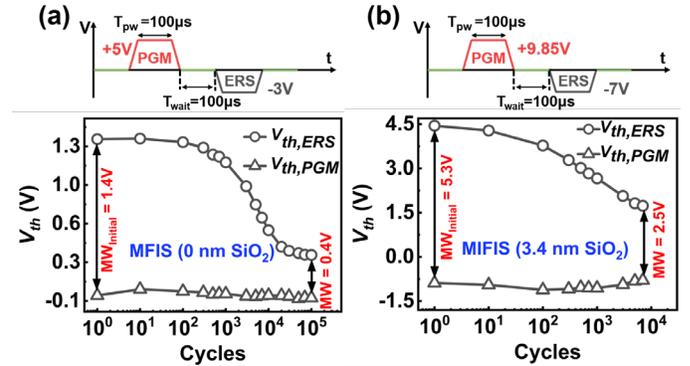

Fig. 10. Endurance characteristics of (a) the MFIS structure and (b) the MIFIS structure with 3.4 nm top SiO$_2$.

of the $V_{th, ERS}$ vs. the top SiO$_2$ thickness curve as shown in Fig. 7(b).

Similarly, we discuss the physical origin of the first stage of the program case. Fig. 8(c) shows the band diagram of the MIFIS during the program operation. During the program operation, the acceptor traps are empty through electrons de-trapping and/or holes trapping regardless of the top SiO$_2$ thickness and show neutral charges. For the SiO$_2$ thickness of less than 1.7 nm, the Fermi energy level of the gate metal TiN is always located below the donor trap energy level $E_{d,1}$ during the program operation. The donor trap energy level $E_{d,1}$ is also empty and shows positive charges. After the program operation, these positive charges are left as shown in Fig. 8(d) and lead to the positive shift of $V_{th}$ after the program ($V_{th, PGM}$) as shown in Fig. 7(b). Moreover, for the SiO$_2$ thickness of less than 1.7 nm, these positive charges determined by donor trap energy level $E_{d,1}$ after the program operation are nearly identical and this results in the first stage of the $V_{th, PGM}$ vs. the top SiO$_2$ thickness curve as shown in Fig. 7(b).

We discuss the physical origin of the second stage of the program case. For the SiO$_2$ thickness of less than 1.7 nm, the donor trap energy level $E_{d,1}$ is empty and the donor trap energy level $E_{d,2}$ is filled by the electrons. For thicker SiO$_2$ thickness of larger than 1.7 nm, however, both the donor trap energy levels $E_{d,2}$ and $E_{d,1}$ are empty because thicker SiO$_2$ induces a larger voltage drop across the whole top SiO$_2$ interlayer and a lower Fermi level of the metal gate with respect to the trap levels as shown in Fig. 8(c). Then more positive charges are left for thicker SiO$_2$ thickness of larger than 1.7 nm, and result in the second stage of the $V_{th, PGM}$ vs. the top SiO$_2$ thickness curve as shown in Fig. 7(b). Finally, the above results lead to a two-stage linear dependence between the maximum MW and the top SiO$_2$ thickness as shown in Fig. 7(a).

The above discussions are verified by the PUND measurement. Fig. 9(a) shows the PUND measurement results. The bulk, source, and drain terminals were shorted during the measurement. As shown in Fig. 9(a), the spontaneous polarization vs. gate voltage ($P$-$V_g$) curve broadens with increasing top SiO$_2$ interlayer thickness. Our measurement results are consistent with [35]. The physical origin of the broadener of the $P$-$V_g$ curve is attributed to the partial voltage across the top SiO$_2$ interlayer caused by the trapped charges at the top SiO$_2$/Hf$_{0.5}$Zr$_{0.5}$O$_2$ interface. We define the distance between the intersections of the $P$-$V_g$ curve and the $V_g$ axis as $V_{g,2Vc}$. Fig. 9(b) exhibits the dependence of $V_{g,2Vc}$ on the top SiO$_2$ interlayer thickness. The dependence, which is similar to that of the maximum MW on the top SiO$_2$ thickness, as shown in Fig. 9(b) further validates the existence of the different interfacial charges $Q_{tp}$ trapped at the top SiO$_2$/Hf$_{0.5}$Zr$_{0.5}$O$_2$ interface.

### C. The endurance and retention characterizations

We study the endurance characteristics of the MIFIS structure. Fig. 10(a) and (b) show the endurance characteristics of the MFIS structure and MIFIS structures, respectively. The MIFIS sample shows an endurance of ~ 10$^4$ cycles when the initial MW is 5.3 V. Additionally, it can be observed that the window stability of the MIFIS structure is superior to that of the MFIS structure. To evaluate the endurance characteristics of the MIFIS structure more reasonably, we investigate the dependence of the endurance characteristics on the initial MW. Fig. 11(a) shows the waveform of endurance characterization measurement. We change the initial MW by iterating through the amplitudes of the program pulse (8-10.5 V). Fig. 11(b) shows the measurement results. The endurance characteristic degrades with a larger initial MW. This is because a larger initial MW means more charges tapping and de-trapping, which accelerates the degradation of the interlayer and leads to poorer endurance characteristics.

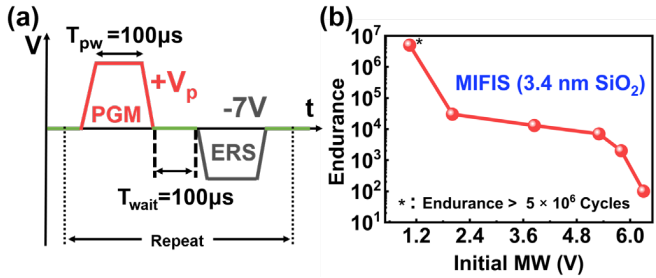

Fig. 11. (a) Waveform of endurance characterization measurement. (b) The dependence of the endurance characteristics on initial MW for MIFIS structure with 3.4 nm top SiO$_2$.

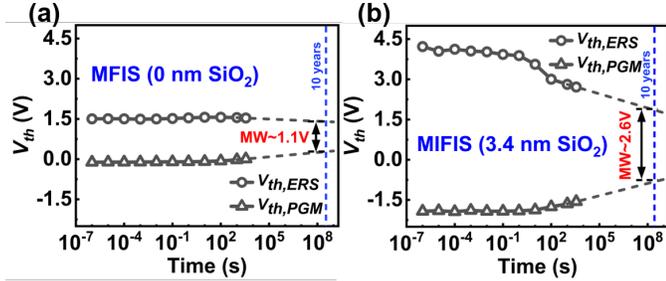

Fig. 12. Retention characteristics measured at room temperature of (a) the MFIS structure and (b) the MIFIS structure with 3.4 nm top SiO$_2$.

TABLE I
Comparisons of recent research with our work

| | Structure | Top IL | MW | Endurance | Retention |
|---|---|---|---|---|---|
| Lee et al. from SK Hynix [21] (2022) | MIFIS | SiO$_2$ | 3.8 V | unknow | 10 years |
| Das et al. from GIT [24] (2023) | MFIFIS | Al$_2$O$_3$ | 7.1 V | unknow | 10 years |
| Suzuki et al. from Kioxia [23] (2023) | MIFIS | unknow | 2.3 V | ~10$^6$ cycles | 10 years |
| Lim et al. from Samsung [22] (2023) | MIFIS | unknow | 3.1 V | ~10$^6$ cycles | 10 years |
| Yoon et al. from SK Hynix [20] (2023) | MIFIS | unknow | 5.04 V | ~3×10$^3$ cycles | 10 years |
| Hu et al. from IMECAS [25] (2024) | MIFIS | Al$_2$O$_3$ | 4.1 V | ~1×10$^4$ cycles | 10 years |
| This work | MIFIS | SiO$_2$ | 5.3 V | ~1×10$^4$ cycles | 10 years |

We study the retention characteristics. Fig. 12 shows the retention characteristics for the two structures under the pulse amplitude of the maximum MW. Both devices have a retention lifetime beyond 10 years.

Table I shows the benchmark of our work. Our work exhibits a larger MW (5.3 V), a better endurance (~ 1 × 10$^4$ cycles), and a retention lifetime beyond 10 years.

IV. CONCLUSIONS

We study the impact of the top SiO$_2$ thickness on the MW. We find that the MW increases with a thicker top SiO$_2$ interlayer. Such an increase exhibits a two-stage linear dependence. We attribute this phenomenon to the presence of different interfacial charges $Q_{tp}$ trapped at the top SiO$_2$/Hf$_{0.5}$Zr$_{0.5}$O$_2$ interface. Furthermore, we also investigate the dependence of the endurance on initial MW, and find that the endurance characteristic degrades with increasing the initial MW. By inserting a 3.4 nm top SiO$_2$ interlayer between the gate metal TiN and the ferroelectric Hf$_{0.5}$Zr$_{0.5}$O$_2$, we achieve the maximum MW of 6.3 V. And our device achieves the endurance ~ 1 × 10$^4$ cycles when the initial MW is ~ 5.3 V and retention lifetime over 10 years.